\documentclass[aps,pre,groupedaddress,amsmath]{revtex4}
\usepackage{graphicx,amssymb}
\usepackage{bm}
\usepackage{amsmath}
\newcommand\beq{\begin{equation}}
\newcommand\eeq{\end{equation}}
\newcommand\beqa{\begin{eqnarray}}
\newcommand\eeqa{\end{eqnarray}}
\newcommand{\dd}{\text{d}}

\newcommand{\nn}{\nonumber\\}

\begin{document}
\title{A granular fluid modeled as a driven system of elastic hard spheres}
\author{Antonio Astillero}
\email{aavivas@unex.es}
\author{Andr\'es Santos}
\email{andres@unex.es}
\homepage{http://www.unex.es/fisteor/vicente/}
\affiliation{Departamento de F\'{\i}sica, Universidad de Extremadura,
E--06071 Badajoz, Spain}
\date{\today}

\begin{abstract}
We explore the possibility of describing the main transport properties of a granular gas by means of a model consisting of elastic hard spheres under the action of a drag force that mimics the inelastic cooling of the granular gas. Direct Monte Carlo simulations of the Boltzmann equation show a good agreement between the results for a gas of inelastic hard spheres and those for a gas of driven elastic hard spheres in the simple shear flow state. This approximate equivalence between both systems is exploited to extend known kinetic models for elastic collisions to the inelastic case.
\end{abstract}

\maketitle

\section{Introduction\label{sec1}}
The simplest model of a granular fluid in the rapid flow regime consists of a gas of (smooth) \textit{inelastic} hard spheres (IHS)  characterized by a constant coefficient of normal restitution $\alpha$ \cite{C90}. The inelasticity of collisions produces  a decrease of the mean kinetic energy (or granular temperature)  with a \textit{cooling rate} $\zeta\propto 1-\alpha^2$. 
Interestingly, 
a cooling effect can also be generated in a gas of  \textit{elastic} hard spheres (EHS) by the application of an effective drag force with a \textit{friction coefficient} $\frac{1}{2}\zeta$. At a macroscopic level of description, the hydrodynamic balance equations of mass, momentum, and energy for the IHS gas are (formally) identical to those for the frictional EHS gas. However, the microscopic dynamics is physically quite different in both systems: in the IHS gas (a) the particles move freely between two successive collisions but (b) each colliding pair loses energy upon collision; in the EHS case (a) the particles lose energy between collisions due to the action of the drag force but (b) energy is conserved by collisions. This implies that during a certain small time step, only the small fraction of colliding particles are responsible for the cooling of the system in the IHS case, whereas all the particles contribute to the cooling in the EHS case. Therefore, there is no reason in principle to expect that the relevant physical properties (e.g., the velocity distribution function) are similar for IHS and frictional EHS under the same conditions.
For instance, in the so-called homogeneous cooling state the solutions to the respective Boltzmann equations for IHS and EHS differ: while the distribution function is a (time-dependent) Gaussian for EHS \cite{GSB90}, deviations from a Gaussian (as exemplified by a nonzero kurtosis and by an overpopulated high energy tail) are present in the case of IHS \cite{vNE98}. 
Notwithstanding this, the differences between the homogeneous solutions for IHS and EHS are not quantitatively important in the domain of thermal velocities and so it is still possible that both systems exhibit comparable departures from equilibrium in inhomogeneous states where transport of momentum and/or energy is the relevant phenomenon. The investigation of this possibility is the main aim of this work.

\section{Model of driven elastic hard spheres\label{sec2}}
The Boltzmann equation for a gas of  inelastic hard spheres (IHS) is \cite{GS87,BDS97}
\beqa
\left( \partial _{t}+{\bf v}\cdot \nabla \right) f({\bf v})&=&\sigma ^{d-1}\int \dd{\bf v}_{1}\int \dd\widehat{\bm{\sigma}}\,\Theta (
{\bf g}\cdot \widehat{\bm{\sigma}})({\bf g}\cdot \widehat{\bm{\sigma}})
\left[ \alpha ^{-2}f({\bf v}')f({\bf v}_{1}')-f({\bf v})f({\bf v}_{1})\right]
\nn
&\equiv&  J^{(\alpha)}[f,f] .
\label{3.2}
\eeqa
In this expression $f({\bf r},{\bf v};t)$ is the one-particle distribution function, $\sigma$ is the diameter of a sphere, $d$ is the dimensionality of the system,
$\Theta $ is the Heaviside step function, $\widehat{
\bm{\sigma}}$ is a unit vector directed along the centers of the two colliding spheres at contact,
$\mathbf{g}=\mathbf{v}-\mathbf{v}_1$ is the relative velocity, and $\alpha$ is the coefficient of normal restitution.
 The precollisional or
restituting velocities ${\bf v}'$ and ${\bf v}_1'$ are
given by 
\begin{equation}
{\bf v}'={\bf v}-\frac{1+\alpha }{2\alpha }({\bf g}\cdot 
\widehat{\bm{\sigma}})\widehat{\bm{\sigma}},\quad {\bf v}_{1}'={\bf v}_{1}+\frac{1+\alpha }{2\alpha }({\bf g}\cdot \widehat{\bm{\sigma}
})\widehat{\bm{\sigma}}.  \label{3.3}
\end{equation}
The collision operator for elastic hard spheres (EHS), $J^{(1)}[f,f]$,  is obtained from eqs.\ (\ref{3.2}) and (\ref{3.3}) by setting $\alpha=1$.

The first $d+2$ moments of the distribution function define the number density $n$, the nonequilibrium flow velocity $\mathbf{u}$ and the \textit{granular} temperature $T$.
The most important properties of $J^{(\alpha)}[f,f]$  are those
that determine the form of the macroscopic balance equations for mass, 
momentum and energy, namely
\begin{equation}
\int \dd{\bf v}
\left\{1,{\bf v},mV^{2}\right\}J^{(\alpha)}[f,f]=
\left\{ 
0,{\bf 0},-{d}nT\zeta\right\}
 ,  \label{3.4}
\end{equation}
where  $m$ is the mass of a particle, ${\bf V}({\bf r},t)\equiv{\bf v}-{\bf u}({\bf r},t)$ is the peculiar velocity and
$\zeta (\mathbf{r},t)$ is the \textit{cooling rate} due to the inelasticity of the collisions. Although it  is a nonlinear functional of the distribution function $f$ \cite{BDS97,BDS99}, a simple \textit{estimate}  $\zeta\approx\zeta_0$ is obtained by replacing the actual distribution function $f$ by the \textit{local equilibrium} distribution $
f_0=n(m/2\pi T)^{d/2}\exp(-mV^2/2T)$. 
The result is \cite{BDS99,BDKS98}
\beq
\zeta_0(\mathbf{r},t)=\nu_0(\mathbf{r},t)\frac{d+2}{4d}(1-\alpha ^{2}),  \quad 
\nu_0\equiv
\frac{8\pi ^{(d-1)/2}\sigma ^{d-1}}{(d+2)\Gamma \left( d/2\right) }n\left( \frac{T}{m}
\right) ^{1/2}
. \label{3.14}
\end{equation}

According to the arguments of section \ref{sec1}, our model consists of the replacement
\beq
J^{(\alpha)}[f,f]\to \beta(\alpha) J^{(1)}[f,f]+\frac{1}{2}\zeta_0 (\alpha)\frac{\partial }{\partial 
{\bf v}}\cdot \left( {\bf V}f\right),
\label{b4}
\eeq
where $\beta(\alpha)$ is a positive constant to be determined. The model (\ref{b4}) complies with the properties (\ref{3.4}), except that the true cooling rate is replaced  by the local equilibrium estimate (\ref{3.14}).
According to eq.\ (\ref{b4}), the gas of \textit{inelastic} hard spheres is replaced by an ``equivalent'' gas of \textit{elastic} hard spheres subjected to the action of a drag force $\mathbf{F}_{\text{drag}}=-(m\zeta_0/2)\mathbf{V}$ proportional to the peculiar velocity with a friction constant that depends on the local density and temperature. This drag force mimics the cooling effect due to dissipative collisions in the underlying granular system.
The parameter $\beta$ accounts for the fact that, in principle, the gas of EHS that more efficiently succeeds in capturing the main properties of the granular gas is made of particles with a diameter $\sigma'=\beta^{1/(d-1)}\sigma$ that does not necessarily coincide with the diameter $\sigma$ of the inelastic spheres. Alternatively, we can view $\beta$ as a correction factor to modify the collision rate of the equivalent system of EHS. A comparison between the transport coefficients of IHS \cite{BDKS98,BC01} and those of the ``equivalent'' EHS \cite{AS03} suggests the choice 
\beq
\beta(\alpha)=\frac{1}{2}(1+\alpha).
\label{beta}
\eeq

\section{Simulations\label{sec3}}
In order to test the model (\ref{b4}) in inhomogeneous states far from equilibrium, we have carried out computer simulations of the Boltzmann equation by means of the DSMC method \cite{B94} in both systems (IHS and EHS) for the simple shear flow problem. In the simple shear flow \cite{C90,GS03}, the gas is enclosed between two infinite parallel plates located at $y=\pm L/2$ and moving with velocities $\pm U/2$ along the $x$-axis. When a particle crosses one of the plates it is reentered through the opposite plate by applying the standard Lees--Edwards boundary conditions \cite{LE72}. This produces a viscous heating effect that tends to increase the temperature of the system, whereas the inelastic cooling (in the IHS system) or the drag force (in the EHS system) tend to decrease the temperature. Eventually, a nonequilibrium steady state (NESS) is reached when both effects cancel each other. If the size of the system is large enough as to avoid clustering effects, the NESS is characterized by uniform density and temperature, and a linear velocity profile $u_x(y)=ay$, where $a=U/L$ is the constant shear rate. 

As a test case, we have considered three-dimensional systems with $L=2.5\lambda$, where $\lambda=(\sqrt{2}\pi\bar{n}\sigma^2)^{-1}$ is the average mean free path of the IHS gas ($\bar{n}$ being the average density), and $U=10 v_0$, where $v_0=\sqrt{2T_0/m}$ is the initial thermal velocity ($T_0$ being the initial temperature). The shear rate is then $a=4 \tau_0^{-1}$, where $\tau_0=\lambda/v_0$ is the initial mean free time of the IHS gas. 
The coefficient of restitution for the IHS gas has been taken as $\alpha=0.9$. As for the EHS gas, its collision rate has been reduced by a factor $\beta=0.95$, in agreement with eq.\ (\ref{beta}). In both cases 
the viscous heating effect dominates during the transient regime until the NESS is reached.
In the simulations we have considered a number $N=10^4$ of simulated particles, a width layer $\delta L=0.05\lambda$ and a time step $\delta t=10^{-3}\tau_0\sqrt{T_0/T}$.

\begin{figure}[h]
\includegraphics[width=\columnwidth]{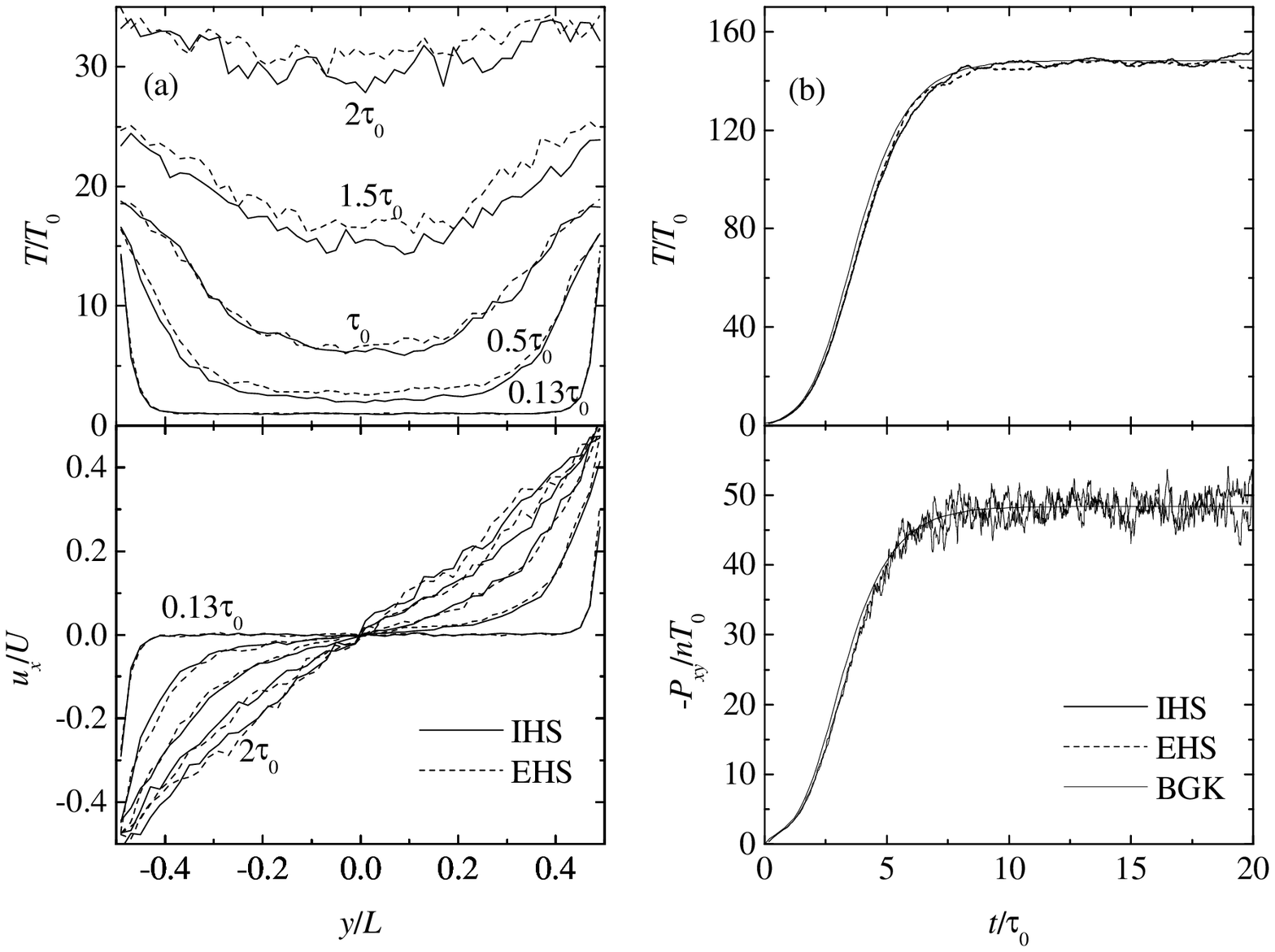}
\caption{(a) Velocity and temperature profiles. (b) Time evolution of the temperature and the shear stress\label{fig1}}
\end{figure}

Figure \ref{fig1}(a) shows the velocity and temperature profiles at times $t/\tau_0=0.13,0.5,1,1.5,2$ for both systems, starting from an initial condition of \textit{total} equilibrium. By time $t=2\tau_0$ (what corresponds to about 2.4 collisions per inelastic particle) the velocity profile is practically linear and the temperature profile is almost uniform. However, the global temperature keeps growing in time until it reaches a stationary value $T\simeq 146 T_0$ for $t\gtrsim 10 \tau_0$ (i.e. after about 48 collisions per inelastic particle). The time evolution of $T/T_0$ and of $-P_{xy}/nT_0$ (where $P_{xy}$ is the shear stress) is shown in fig.\ \ref{fig1}(b) for an initial condition of \textit{local} equilibrium, so that the system is initially prepared with a linear velocity profile. The NESS value of the shear stress represents an effective shear viscosity about 14\% smaller than the Navier--Stokes value corresponding to $\alpha=0.9$.
Figure \ref{fig1}, along with a more comprehensive comparison that will be reported elsewhere \cite{AS03}, shows  that the equivalent EHS system succeeds in capturing the main nonequilibrium transport properties of the underlying IHS system.

\section{Kinetic modeling\label{sec4}}
\subsection{BGK and ES models}
The mapping IHS$\to$EHS allows one to take advantage of the existence of simple kinetic models for EHS to extend them straightforwardly to IHS. For instance, consider the so-called ellipsoidal statistical (ES) model \cite{GS03,Cer90}
\beq
J^{(1)}[f,f]\to -\nu_0(1-\epsilon)(f-f_\epsilon),
\label{7.8}
\eeq
where
\begin{equation}
f_\epsilon(\mathbf{v})
=n\left(\frac{mn}{2\pi}\right)^{d/2} \left(\det \mathsf{R}\right)^{-1/2}\exp\left(-\frac{mn}{2}\mathsf{R}^{-1}:\mathbf{V}\mathbf{V}\right),\quad \mathsf{R}=\frac{1}{1-\epsilon}\left(p\mathsf{I}-\epsilon\mathsf{P}\right),
\label{6.5:n2}
\end{equation}
$\mathsf{P}$ being the pressure tensor. The parameter $\epsilon\leq d^{-1}$ can be freely chosen. The choice $\epsilon=d^{-1}$ makes the ES model reproduce the correct  value $\text{Pr}=(d-1)/d$ of the Prandtl number.
The simplest choice, however, is $\epsilon=0$, in which case $\mathsf{R}=p\mathsf{I}$, the reference function $f_\epsilon=f_0$ becomes the local equilibrium distribution and the ES model reduces to the well-known BGK model.

In the spirit of (\ref{b4}), the extension of the ES model to IHS is
\beq
J^{(\alpha)}[f,f]\to -\beta(\alpha)\nu_0(1-\epsilon)(f-f_\epsilon)+\frac{1}{2}\zeta_0 (\alpha)\frac{\partial }{\partial 
{\bf v}}\cdot \left( {\bf V}f\right).
\label{7.9}
\eeq
In particular, setting $\epsilon=0$ we get a simplified version of the BGK-like model for IHS that 
was proposed in ref.\ \cite{BDS99}. This BGK model is easy to solve for the simple shear flow problem considered in section \ref{sec3}. The thin solid lines in fig.\ \ref{fig1}(b) represent the evolution of the temperature and the shear stress according to such a solution. As can be observed, the BGK solution exhibits an excellent agreement with the DSMC results for both IHS and EHS.

\subsection{Mixtures}
The same idea behind (\ref{b4}) can be extended to a multi-component granular gas \cite{AS03}. In the special case where all the species have the same flow velocity ($\mathbf{u}_i=\mathbf{u}$), our model becomes
\beq
J_{ij}^{(\alpha_{ij})}[f_i,f_j]\to \beta_{ij}(\alpha_{ij}) J_{ij}^{(1)}[f_i,f_j]+\frac{1}{2}\zeta_{ij}(\alpha_{ij})\frac{\partial }{\partial 
{\bf v}}\cdot \left[\left( {\bf v }-\mathbf{u}_i\right)f_i\right]
\label{5.5}
\eeq
with
\beq
\beta_{ij}=\frac{1+\alpha_{ij}}{2}, \quad
\zeta_{ij}=\frac{\sqrt{2}\pi^{(d-1)/2}}{\Gamma(1+d/2)}n_j\mu_{ji}^2\sigma_{ij}^{d-1}\left(\frac{T_i}{m_i}\right)^{1/2}\left(1+\frac{m_iT_j}{m_jT_i}\right)^{3/2}\left(1-\alpha_{ij}^2\right),
\label{5.13}
\eeq
where $T_i$ is the granular temperature of species $i$ and $\mu_{ji}\equiv (1+m_i/m_j)^{-1}$.
The model (\ref{5.5}) preserves the first $d+2$ collision integrals of IHS in the leading Sonine approximation with $\mathbf{u}_i=\mathbf{u}$.
The important point is that the approximation (\ref{5.5}) allows one to transfer any given kinetic model 
\beq
J_{ij}^{(1)}[f_i,f_j]\to K_{ij}^{(1)}
\label{5.22}
\eeq
for \textit{elastic} mixtures \cite{GS03} into an equivalent model for \textit{inelastic} mixtures:
\beq
J_{ij}^{(\alpha_{ij})}[f_i,f_j]\to K_{ij}^{(\alpha_{ij})}= \frac{1+\alpha_{ij}}{2}K_{ij}^{(1)}
+\frac{\zeta_{ij}}{2}\frac{\partial }{\partial 
{\bf v}}\cdot \left[\left( {\bf v }-\mathbf{u}_{i}\right)f_i\right].
\label{5.23}
\eeq

\section{Conclusion\label{sec5}}
In summary, we have shown that the nonequilibrium transport properties of a Boltzmann gas of inelastic hard spheres can be satisfactorily captured by an equivalent gas of elastic hard spheres driven by a dissipative drag force $\mathbf{F}_{\text{drag}}=-(m\zeta_0/2)\mathbf{V}$, where $\zeta_0(\alpha)$ is the (local equilibrium) cooling rate. Besides, the elastic particles must reduce their collision rate by a factor $\beta(\alpha)\approx \frac{1}{2}(1+\alpha)$ in order to ``disguise'' as a granular gas. While the ``equivalent'' system of EHS does not retain finer details of the true IHS gas (e.g., high energy tails, velocity correlations, \ldots), it is able to account for those phenomena (e.g., inelastic clustering) that can be described at a hydrodynamic level.
Finally, we have exploited the possibility of reverting the mapping IHS$\to$EHS to construct kinetic models for granular gases as  natural extensions of known kinetic models originally proposed for elastic particles.

\acknowledgments
This work has been supported by the
Ministerio de
Ciencia y Tecnolog\'{\i}a
 (Spain) through grant No.\ BFM2001-0718. A.A. is grateful to the Fundaci\'on Ram\'on Areces for a predoctoral fellowship.

\end{document}